\documentclass[a4paper,11pt]{article}
\usepackage{pos}
\usepackage{slashed}
\usepackage{graphicx}

\newcounter{notecount}

\newcommand{\be}{\begin{equation}}
\newcommand{\ee}{\end{equation}}

\newcommand{\lt}{\left}
\newcommand{\rt}{\right}
\newcommand{\del}{\partial}

\newcommand{\non}{\nonumber \\}

\newcommand{\fn}{\footnote}

\newcommand{\MSb}{\overline{\rm MS}}

\newcommand{\mbar}{\overline{m}}

\title{Determining the Quark Mass with the Gradient Flow}
\ShortTitle{Quark Mass with Gradient Flow}

\author*[a]{Hiromasa Takaura}
\author[b]{Robert Harlander}
\author[c,d]{Fabian Lange}

\affiliation[a]{Center for Gravitational Physics and Quantum Information, Yukawa Institute for Theoretical Physics, Kyoto University,\\
Kyoto 606-8502, Japan}

\affiliation[b]{Institute for Theoretical Particle Physics and Cosmology, RWTH Aachen University,\\
  52056 Aachen, Germany}

\affiliation[c]{Physik-Institut, Universit\"at Z\"urich, \\
Winterthurerstrasse 190, 8057 Z\"urich, Switzerland}

\affiliation[d]{PSI Center for Neutron and Muon Sciences,\\
Forschungsstrasse 111, 5232 Villigen PSI, Switzerland}

\emailAdd{hiromasa.takaura@yukawa.kyoto-u.ac.jp}

\abstract{We propose a new method to determine the quark mass
by using bilinear operators of the flowed quark field defined within the gradient-flow formalism.
This method enables the quark mass determination through a comparison of perturbative calculations with lattice data.
The gauge-invariant nature of the observable should allow clear control over perturbative errors.
At the same time, the gradient flow suppresses the noise in the lattice measurements of the observable, which simply consists of one-point functions.
Concerning the perturbative input in this framework, we study the mass dependence
of the flowed quark condensate $\langle \bar{\chi}(t,x) \chi(t,x) \rangle$ at the two-loop level. For this purpose,
we develop a novel approach for expanding massive gradient-flow integrals in the limit
of small and large $(m^2t)$. We also present a fully numerical result which
includes the full mass dependence.
}

\FullConference{The 41st International Symposium on Lattice Field Theory (LATTICE2024)\\
 28 July - 3 August 2024\\
Liverpool, UK\\}

\begin{document}
\maketitle

\section{Introduction}

Quark masses are fundamental parameters in QCD, and their accurate knowledge is essential for precision physics.
For instance, uncertainties in the heavy-quark $\overline{\rm MS}$ masses contribute to theoretical errors in
the corresponding Yukawa couplings to the Higgs field. Their determination is
crucial for testing the
mechanism of electroweak symmetry breaking.
The precision of heavy-quark masses also plays an important role in flavor physics.

Heavy-quark ($b, c$) masses have already been determined by many lattice investigations and we refer to Ref.~\cite{FlavourLatticeAveragingGroupFLAG:2024oxs} for an overview of the available results.
The main tools to relate the lattice to the customary modified minimal subtraction $\big(\overline{\rm MS}\big)$ results used in these studies are the regularization-independent momentum subtraction (RI-MOM) scheme~\cite{Martinelli:1994ty},
its variant regularization-independent symmetric momentum subtraction (RI-SMOM) scheme~\cite{Sturm:2009kb}, the quark current (or current-like) correlator method~\cite{HPQCD:2008kxl}, or the minimal renormalon-subtracted (MRS) scheme~\cite{Brambilla:2017hcq}.

In these proceedings, we propose an alternative way to determine the heavy-quark mass
on the lattice based on the gradient-flow formalism~\cite{Narayanan:2006rf,Luscher:2010iy,Luscher:2013cpa}.
Its potential to extract $\alpha_s(M_Z^2)$ from lattice data has recently been explored in Refs.\,\cite{Hasenfratz:2023bok,Wong:2023jvr,Schierholz:2024lge}.
For the purpose of the heavy-quark masses, we consider the following ratio of flowed quark condensates:
\begin{equation}
\mathring{S}(t, \overline{m}_1, \overline{m}_2) \equiv \frac{\langle \bar{\chi}_1(t,x) \chi_1(t,x) \rangle}{\langle  \bar{\chi}_2(t,x) \overleftrightarrow{\slashed{D}} \chi_2(t,x) \rangle} , \label{eq1}
\end{equation}
where $\chi_i(t,x)$ and $\bar{\chi}_i(t,x)$ denote the flowed fermion fields defined by the flow equations \cite{Narayanan:2006rf,Luscher:2010iy,Luscher:2013cpa}
\begin{align}
\partial_t \chi_i(t,x) &=(D_{\mu} D_{\mu}-\alpha_0 \partial_{\mu} B_{\mu}(t,x) )\chi_i(t,x) , \non
\partial_t \bar{\chi}_i(t,x) &=\bar{\chi}_i(t,x)(\overleftarrow{D}_{\mu} \overleftarrow{D}_{\mu}+\alpha_0 \partial_{\mu} B_{\mu}(t,x) ) , \non
\del_t B_{\mu}(t,x)&=D_{\nu} G_{\nu \mu}(t,x)+\alpha_0 D_{\mu} \del_{\nu} B_{\nu}(t,x) ,
\end{align}
and the boundary conditions $\chi_i(t=0,x)=\psi_i(x)$,
$\bar{\chi}_i(t=0,x)=\bar{\psi}_i(x)$, $B_{\mu}(t=0,x)=A_{\mu}(x)$.
$\alpha_0$ is a gauge fixing parameter which cancels in physical
quantities; throughout our calculation, we choose $\alpha_0=1$.
The renormalized mass of the quark field $\psi_i$ in the $\MSb$ scheme is denoted by $\overline{m}_i$.  Even though both
numerator and denominator in Eq.~\eqref{eq1} are separately ultraviolet divergent, the
ratio is finite~\cite{Makino:2014taa} because the divergences only affect the
wave function renormalization factor \cite{Luscher:2013cpa}. Thus, the
continuum limit in a lattice determination of $\mathring{S}(t, \overline{m}_1,
\mbar_2)$ exists.  On the other hand, one can calculate $\mathring{S}(t,
\overline{m}_1, \mbar_2)$ in perturbation theory, where the result is
expressed as a function of $t$, $\alpha_s$, $\mbar_1$, and $\mbar_2$.  By
matching the perturbative and the lattice results, one should thus be able to
determine the $\MSb$ mass(es).

The minimal uncertainty achievable within perturbation theory for an observable is determined by nonperturbative effects.
Since $\mathring{S}(t, \overline{m}_1, \mbar_2)$ of Eq.~\eqref{eq1} is a gauge invariant quantity, the leading
non-perturbative contributions are given by the dimension-four condensates
such as the
gluon condensate $\langle  F^{a}_ {\mu \nu} F^{a}_ {\mu \nu} \rangle $, because gauge and Lorentz invariance
prohibit (parametrically larger) dimension-two condensates.
This argument does not apply to gauge-dependent methods
such as the RI-(S)MOM scheme.
Other advantages of our proposal are related to the fact that
$\mathring{S}(t,\mbar_1, \mbar_2)$ is given by one-point functions, and the fact that the noise in
its lattice determination is suppressed by the gradient flow.

The perturbative result for $\mathring{S}(t, \mbar,  \mbar)$ is known at next-to-next-to-leading order QCD in the limit of a small
quark mass $\mbar$: Concerning the numerator in Eq.\,\eqref{eq1}, it vanishes in the strictly
massless limit. The linear mass term, on the other hand, was calculated at
$\mathcal{O}(\alpha_s)$ in Ref.~\cite{Makino:2014taa}\footnote{To be more precise, it was presented in the arXiv version v2 of Ref.~\cite{Makino:2014taa}.}
and at $\mathcal{O}(\alpha^2_s)$ in Ref.~\cite{Artz:2019bpr}.
The massless limit of the denominator of $\mathring{S}(t, \mbar,  \mbar)$ was calculated
at $\mathcal{O}(\alpha_s)$ in Ref.~\cite{Makino:2014taa} and at $\mathcal{O}(\alpha^2_s)$ in Refs.~\cite{Harlander:2018zpi,Artz:2019bpr}.

However, the size of a dimensionless combination $\mbar^2 t$ can span a wide
range in the heavy-quark cases.
Let us impose
$a^2 \ll 8 t \ll L^2$ and $8 t \ll \Lambda_{\rm QCD}^{-2}$ for the range of
the flow time $t$, where $a^{-1}=4$~GeV is assumed as a reference and $\Lambda_{\rm QCD} \sim 0.3$~GeV.
The first condition is necessary for reasonable lattice simulations of Eq.~\eqref{eq1},
and the second one is imposed from perturbativity.
Then,
$0.1 \ll 8 \mbar_c^2 t \ll 20$ in the charm-quark case and $1.0 \ll 8 \mbar_b^2 t \ll 200$ in the bottom-quark case.
Therefore, the known results, which are valid for $\mbar^2 t \ll 1$, are not always sufficient.

In this paper, we investigate the mass dependence of the numerator of Eq.~\eqref{eq1},
i.e., the quark condensate $\langle \bar{\chi}_f \chi_f \rangle$ with $\mbar_f=\mbar$, at
$\mathcal{O}(\alpha_s)$ in perturbation theory,
as a first step toward the mass determination in this framework.
We present the first two terms of the small-$\mbar^2 t$ expansion, and
the first three terms of the large-$\mbar^2 t$ expansion.
Additionally, we present the full mass dependence, which is valid for a wide
range of $\mbar^2 t$, based on a numerical computation.

\section{Perturbative computation}

For the foundations of the perturbative approach to the QCD gradient flow we refer to Refs.~\cite{Luscher:2011bx,Artz:2019bpr}.
In the computation of $\langle \bar{\chi}(t,x) \chi(t,x) \rangle$ at $\mathcal{O}(\alpha_s)$, we find eight scalar integrals.
Examples include
\begin{align}
&\int_0^t \mathrm{d}s \int_0^s \mathrm{d}s' \int_{p,k} \frac{m p^2}{k^2 (p^2+m^2)} e^{-(2t-s+s') p^2-(s+s') k^2-(s-s') (k-p)^2} ,\non
&\int_0^t \mathrm{d}s \int_{p,k} \frac{m}{k^2 (p^2+m^2)} e^{-(2t-s) p^2-s k^2-s(k-p)^2} , \non
&\int_0^t \mathrm{d}s \int_0^t \mathrm{d}s' \int_{p,k} \frac{m (p-k)^2}{k^2((p-k)^2+m^2)} e^{-(2t-s-s')p^2-(s+s') k^2-(s+s') (k-p)^2} ,
\end{align}
where $\int_{p,k} \equiv \int \frac{\mathrm{d}^d p}{(2 \pi)^d} \frac{\mathrm{d}^d k}{(2 \pi)^d}$ with $d=4-2 \epsilon$ and $m$ is the bare mass.
Flow-time integrals appear in addition to the loop-momentum integrals.

We evaluate these integrals in an expansion in either $m^2 t$ or $1/(m^2 t)$.
The standard technique to expand loop integrals under a certain hierarchy is known as ``expansion by regions''  \cite{Beneke:1997zp}.
Its application to the gradient flow formalism has been described in
Ref.~\cite{Harlander:2021esn}. It was pointed out though that, in the
large-$m^2t$ limit, this method is not straightforwardly applicable, because
the flow-time integration extends over hard and soft regions simultaneously.
In the present study, we pursue a different strategy which does not suffer
from this problem.
Given an integral $I(m^2,t)$, we perform the Laplace transform
\be
\tilde{I}(v,t) \equiv \int_0^{\infty} \mathrm{d} (m^2) (m^2)^{-v-1} I(m^2,t) .
\ee
In the present case, $\tilde{I}(v,t) \propto t^{v-d+5/2}$ as seen from dimensional analysis.
The inverse transform is given by
\be
I(m^2,t)=\frac{1}{2 \pi \mathrm{i}} \int_{-\mathrm{i} \infty}^{\mathrm{i} \infty} \mathrm{d} v \tilde{I}(v,t) (m^2)^v . \label{Laplace}
\ee
For $m^2 t \ll 1$, we close the integration contour of the inverse transform in the right half of the $v$-plane,
considering $(m^2 t)^v=e^{v \log{(m^2 t)}} \to 0$ as $v \to + \infty$.
We then obtain the small-$(m^2 t)$ expansion by
\be
I(m^2,t;d)=-\sum_{v_0>0} {\rm Res}  [\tilde{I}(v,t;d) (m^2)^v]|_{v=v_0} , \label{smallmassexpformula}
\ee
where $v_0$ denotes positive singularities of $\tilde{I}(v,t) $.
For $m^2 t \gg 1$, in turn, we obtain the large-$(m^2 t)$ expansion by
\be
I(m^2,t;d)=\sum_{v_0<0} {\rm Res}  [\tilde{I}(v,t;d) (m^2)^v]|_{v=v_0} ,  \label{largemassexpformula}
\ee
where $v_0$ denotes negative singularities of $\tilde{I}(v,t)$.
To obtain these expansions, we need to calculate the singularities of the integrand in Eq.~\eqref{Laplace}.
This approach is analogous to the ideas found in Refs.~\cite{Neubert:1994vb,Kitano:2022gzy}.
Further details will be explained in a future publication.

We obtain
\begin{align}
&\langle [\bar{\chi}(t,x) \chi(t,x) ]_R \rangle \non
&=-\frac{N_c}{8 \pi^2} \frac{\overline{m}}{t}  \non
&\times  \bigg\{ 1+0.759581C_F \alpha_s(\mu^2)   \non
&\quad{}+\overline{m}^2 t \big[2 \log{2}+2 \gamma_E+2 \log{(\overline{m}^2 t)}+C_F \alpha_s(\mu^2) (2.3334-2.16804 \log{(\overline{m}^2/\mu^2}) \non
&\qquad{}\qquad{}+\log{(\overline{m}^2 t)}(2.22817-0.95493 \log{(\overline{m}^2/\mu^2}) )\big] \non
&\quad{}+\mathcal{O} ((\overline{m}^2 t)^2) \bigg\} , \label{smallmassres}
\end{align}
and
\begin{align}
&\langle [\bar{\chi}(t,x) \chi(t,x) ]_R \rangle \non
&=-\frac{N_c}{16 \pi^2 t^2 \overline{m}}  \non
&\times  \bigg\{ 1+\frac{C_F \alpha_s(\mu^2)}{4 \pi} (-2-3 \gamma_E-9 \log{2}+9 \log{3}-3 \log{(\overline{m}^2 t)}+6 \log{(\overline{m}^2/\mu^2}) ) \non
&+\frac{1}{\overline{m}^2 t} \lt[-1+\frac{C_F \alpha_s(\mu^2)}{16 \pi} (22+21 \gamma_E+63 \log{2}-51 \log{3}+21 \log{(\overline{m}^2 t)}-48 \log{(\overline{m}^2/\mu^2}) )\rt] \non
&+\frac{1}{(\overline{m}^2 t)^2}  \lt[\frac{3}{2}+\frac{C_F \alpha_s(\mu^2)}{64 \pi}
( -3 (68+57 \gamma_E)+513 \log{2}+381 \log{3}-171 \log{(\overline{m}^2 t)}+ 432\log{(\overline{m}^2/\mu^2}) )\rt] \non
&+\mathcal{O} (\frac{1}{(\overline{m}^2 t)^3}) \bigg\} . \label{largemassres}
\end{align}
Here we give the finite result by renormalizing the flowed field as $[\bar{\chi}(t,x) \chi(t,x)]_R \equiv Z_{\chi}  \bar{\chi}(t,x) \chi(t,x)$, in the $\MSb$ scheme,
where $Z_{\chi}=1+\frac{\alpha_s(\mu^2)}{4 \pi} \frac{6 C_F}{2 \epsilon}+\mathcal{O}(\alpha_s^2)$ \cite{Luscher:2013cpa}.
We also renormalize the bare mass as $m=Z_m^{\overline{\rm MS}} \overline{m}$ with
$Z_m^{\overline{\rm MS}}=1-\frac{\alpha_s}{4 \pi} \frac{6 C_F}{2 \epsilon} +\mathcal{O}(\alpha_s^2)$.
In the above results, $\mbar=\mbar(\mu^2)$.
All the $\mathcal{O}(\alpha_s^0)$ terms were calculated in Ref.~\cite{Harlander:2021esn}.\fn{%
Typos in the published version of Ref.~\cite{Harlander:2021esn} are corrected in the latest arXiv version.}
The next-to-leading term in $\mbar^2 t$ at $\mathcal{O}(\alpha_s)$ of
Eq.~\eqref{smallmassres} is new, whereas in Eq.~\eqref{largemassres},
all the $\mathcal{O}(\alpha_s)$ terms are new.

Besides these expansions, we numerically evaluate the scalar integrals using \texttt{ftint} \cite{Harlander:2024vmn}.
In this calculation, we do not need to assume large or small $\mbar^2 t$
and can obtain the $\mbar^2 t$-dependence of $\langle [\bar{\chi}(t,x) \chi(t,x) ]_R \rangle$ in a wide range.

\begin{figure}[tb]
\begin{minipage}{0.5\hsize}
\begin{center}
\includegraphics[width=0.95\textwidth]{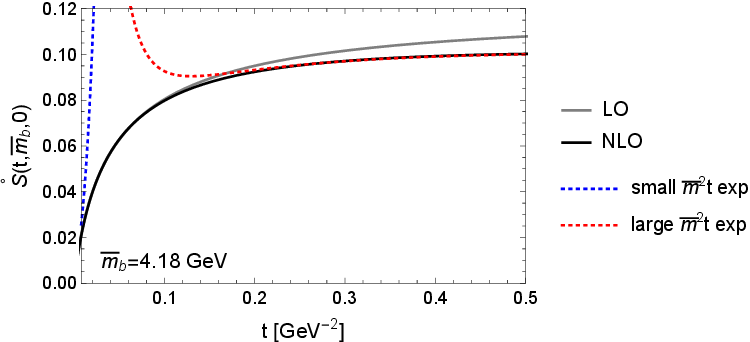}
\end{center}
\end{minipage}
\begin{minipage}{0.5\hsize}
\begin{center}
\includegraphics[width=0.95\textwidth]{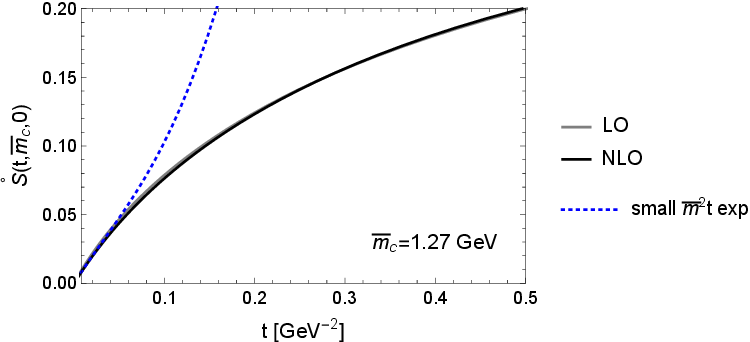}
\end{center}
\end{minipage}
\caption{$\mathring{S}(t, \mbar, 0)$ as a function of the flow time $t$ for the bottom-quark case ($\mbar=\mbar_b$) (left) and
the charm-quark case ($\mbar=\mbar_c$) (right). The gray line shows the $\mathcal{O}(\alpha_s^0)$-result with full mass dependence,
and the black line shows the $\mathcal{O}(\alpha_s)$-result with full mass dependence obtained by our numerical study.
The small-$(\mbar^2 t)$ expansion and the large-$(\mbar^2 t)$ expansion to $\mathcal{O}(\alpha_s)$ are also shown.
(The large-$(\mbar^2 t)$ expansion is not relevant to the charm quark case.)}
\label{fig}
\end{figure}

Using these results, we show the perturbative result of $\mathring{S}(t, \mbar, 0)$ in Fig.~\ref{fig} as a function of the flow time $t$.
For the denominator of $\mathring{S}(t, \mbar, 0)$, we use the known massless result to $\mathcal{O}(\alpha^1_s)$.
In these figures, we express our result in terms of $\mbar(\mbar)$ and $\alpha_s(\mu_t)$,
where $\mu_t=\frac{1}{\sqrt{2te^{\gamma_E}}}$~\cite{Harlander:2018zpi}.
The running coupling is evaluated for four (five) active quark flavors in the
case of the charm (bottom) quark.
As input parameters, we adopt $\alpha_s(M_Z^2)=0.1179$,
$\mbar_b(\mbar_b)=4.18$~GeV and $\mbar_c(\mbar_c)=1.27$~GeV.
The figures indicate the validity of the small- and
large-$m^2t$ expansions and provide a convincing verification of our expansion method as well,
by the agreement of these approximations with the
numerical results.

Assuming that the corresponding lattice data are available,
one can perform a mass determination at the level of the precision of
$\mathcal{O}(\alpha_s)$ with this perturbative result.

\section{Summary and outlook}

The gradient-flow formalism has proven useful already in many different
contexts. In
this contribution, we proposed a new method to determine heavy-quark masses by
utilizing quark bilinear operators defined in the gradient-flow formalism.
Eq.~\eqref{eq1} is manifestly gauge invariant and simply a combination of
one-point functions defined in the gradient flow.  These features have the
potential to contribute to a precise determination of the quark masses.

To perform the mass determination in this framework, one needs a perturbative result to match with lattice data.
We calculated the quark condensate $\langle \bar{\chi}(t,x) \chi(t,x) \rangle$ at $\mathcal{O}(\alpha_s)$,
keeping its quark mass dependence in an expansion in either $\mbar^2t$ or $1/(\mbar^2t)$,
and by a numerical computation valid for a wide range of $\mbar^2t$.

While the first few terms of the small- and large-$\mbar^2 t$ expansions were
given here,
we will extend these expansions further and present them in a future paper.
We will also study the mass dependence of $\langle  \bar{\chi}(t,x) \overleftrightarrow{\slashed{D}} \chi(t,x) \rangle$.
Furthermore, one may study other observables such as $1-\langle  \bar{\chi}_1(t,x) \overleftrightarrow{\slashed{D}} \chi_1(t,x) \rangle/\langle  \bar{\chi}_2(t,x) \overleftrightarrow{\slashed{D}} \chi_2(t,x) \rangle$,
where $\chi_1$ is a heavy and $\chi_2$ is a massless quark.
This observable exhibits the heavy-quark mass dependence of $\mathcal{O}(\mbar^2)$ for small $\mbar^2 t$
and is therefore expected to be highly sensitive to the quark mass,
while, for instance, $\mathring{S}(t, \mbar, 0)$ and $\mathring{S}(t, \mbar, \mbar)$ exhibit
the $\mathcal{O}(\mbar^1)$ dependence.

Although the determination of heavy-quark masses motivated us to study the mass dependence for a wide range of $\mbar^2 t$,
this framework should be applicable to the determination of light-quark masses as well.

For a precise determination, it is crucial to calculate higher orders in $\alpha_s$.
A perturbative calculation at $\mathcal{O}(\alpha_s^2)$ with mass dependence might be feasible.
It is also theoretically important to address the structure of the small-flow-time expansion for these observables,
as understanding this structure can help reduce or properly account for theoretical uncertainties.
Lastly, the precision that can be achieved on the lattice in practice is crucial and remains to be investigated.

\acknowledgments HT thanks Masakiyo Kitazawa for fruitful discussion.
HT is Yukawa Research Fellow supported by the Yukawa Memorial Foundation. The work of HT
was also supported by JSPS Grants-in-Aid for Scientific Research with Numbers
JP19K14711 and JP23K13110.
The work of RH was supported by the Deutsche
Forschungsgemeinschaft (DFG, German Research Foundation) under grant
460791904. The work of FL was supported by the Swiss National Science Foundation (SNSF) under contract \href{https://data.snf.ch/grants/grant/211209}{TMSGI2\_211209}.

\bibliographystyle{utphys}
\bibliography{GradientFlow.bib}

\end{document}